\documentclass[a4paper, 10pt,twocolumn,twoside]{IEEEtran}
\usepackage{amssymb,amsmath,amsfonts}
\usepackage{color}
\usepackage{mathtools}
\usepackage{mathrsfs}
\usepackage{bm}
\usepackage{graphicx}		
\usepackage{epstopdf}		
\usepackage[font={small}]{caption}
\usepackage{caption}
\usepackage{subcaption}
\usepackage{multicol}
\usepackage{enumerate}
\usepackage{units}
\usepackage[numbers]{natbib}
\usepackage{algorithm}
\usepackage{color}

\usepackage{algorithm}
\usepackage{algpseudocode}
\usepackage{cuted}
\usepackage{epsfig}

  \usepackage{amsthm}

\makeatletter
\def\BState{\State\hskip-\ALG@thistlm}
\makeatother

\usepackage{flushend}

\DeclareMathOperator*{\argmax}{arg\,max}
\newtheorem{prep}{Proposition}

\newtheorem{prob}{Problem}
\newtheorem{rema}{Remark}

\DeclarePairedDelimiter\norm{\lVert}{\rVert}

\begin{document}
\title{Energy Efficient Delay Sensitive Optimization in SWIPT-MIMO}
 \author{Hadi Saki, Tong Peng, M. Shikh Bahae \\
\IEEEauthorblockA{Institute of Telecommunications, King's College London, London, United Kingdom\\
Email: \{hadi.saki,tong.peng,m.sbahaei\}@kcl.ac.uk}}
\vspace{-3em}
\thispagestyle{empty}
\maketitle
\pagestyle{empty}
\begin{abstract}
 In this paper, we consider joint antenna selection and optimal beamforming for energy efficient delay minimization. We assume multiple-input multi-output (MIMO) system with full duplex simultaneous wireless information and power transfer (FD-SWIPT) where each sensor is equipped with a power splitting (PS) system and can simultaneously receive both energy and information from the aggregator (AGG). 
We show that the antenna selection and beamforming power control policies are adaptive to the energy state information (ESI), the queue state information (QSI) and the channel state information (CSI). We develop an analytical framework for energy efficient delay-optimal control problem based on the theory of infinite horizon partially observable Markov decision process (POMDP). The infinite-horizon POMDP problem is transformed into an equivalent value Bellman program and solved by
near-optimal point-based Heuristic Search Value Iteration (PB-HSVI) method under specific standard conditions. 
The proposed solution outcome is a set of sub-optimal antenna selection and beamforming control policies. Simulation results reveal an effective trade-off between the contradictory objectives (i.e. delay and power consumption) and show the enhancement in delay by using FD-SWIPT systems in comparison to Half Duplex (HD)-SWIPT systems.
\end{abstract}
\section{Introduction}
\thispagestyle{empty}
Real-time and reliable data transmission is an essential requirement for next-generation autonomous platforms where power consumption, latency, and delay are important parameters . 
In terms of communication systems, research has addressed these requirements in several ways including cross layer design \cite{Shadmand2010,Nehra2010}, energy efficient communication technologies \cite{Olfat,Mahyari2015}, and improved reliable and fast communication systems. 
Simultaneous wireless information and power transfer  (SWIPT) technology is utilized to transport both energy and information. For example, the beamforming optimization of the single user full duplex (FD) \cite{Towhidlou2016,Naslcheraghi2017} SWIPT communication systems was studied in \cite{Chalise:}. The work in \cite{Huang2013}, studied the Lyapunov optimization framework in energy harvesting network to derive an efficient energy management algorithm.
Authors in \cite{Zhang2013}  characterized the energy-rate region of WPT MIMO network. Energy efficiency (EE) optimization problem in SWIPT MIMO broadcast channel was studied in \cite{Tang2017}. 
Authors in \cite{7906591} also investigated the energy-efficient joint beamforming and antenna selection of massive mimo transmission with imperfect CSI. 
In-band FD (IBFD) massive mimo spectral efficiency was studied in \cite{7873882}. In this work, the lower and upper bound of IBFD capacity was derived. Authors in \cite{Jang:} proposed a near optimal antenna selection on two-way FD mimo communication systems. The sub-optimal power allocation and beamforming problem to optimize the max-min weighted SINR problem for multiple half duplex downlink and uplink users and full duplex multi user mimo base station was investigated in \cite{Jiang:}. However,  the lower layer characteristics like energy efficiency and throughput performance are mainly considered in the most of the works,  and the dynamics of data queue state information (QSI) and bursty video data arrival are disregarded.  
Only a few works have addressed the delay sensitive resource management policies \cite{Wang:,Cui:a}. 
The traditional CSI-based beamforming and antenna selection policy usually favor the user with the lowest interference (i.e. the user at the center of the cluster) while the SUs with higher interference (SUs at the cluster edge) are usually neglected. This may result in a severe delay of SUs at the edge of the cluster,  consequently,  severe average delay of the network. By contrast, the QSI, ESI and CSI aware control policy will adaptively favour beamforming and antenna selection policy based on the users QSI, ESI and CSI to address the SUs battery state,  data flow urgency and the channels state condition. Consequently, it will provide a better average delay. However, jointly considering the energy state, the physical layer and queuing delay performance management in MIMO wireless sensor network is not a trivial problem as it will require queuing theory (to perform the energy and data queuing dynamic models) and also involves information theory.
\subsection{Contributions}
In this work, we investigate the delay sensitive problem in an IBFD SWIPT MIMO system where the Sensor Users (SUs') harvested energy, beamforming as well as the data queue are considered. Our objective is to minimize the SUs delay under minimum average power and average rate constraints, by optimally selecting the active antenna set policy and optimizing the beamforming. The considered POMDP
optimization problems are extremely challenging to solve, considering that they are non-convex infinite integer problems. To be able to solve these problems, we developed a two-layer method where the beamforming procedure is divided from the antenna selection operation. For a fixed policy of active antenna set, a sub-optimal upper-bound beamforming method based on the point-based Heuristic Search Value Iteration (PB-HSVI) method is developed  (first-layer).  Specifically, the corresponding belief states, observation states, cost function and value function need to be updated through the increased uncertainty of a  reachable belief states and apply the piece-wise linear and convex value function optimization for each iteration to reduce the complexity. In the second layer, we developed a low-complexity iterative beamforming method to mitigate the inter and intra interference.
To further reduce the computational complexity, we introduced the Stochastic Simulation by Explorative Action heuristic (SSEA) algorithm for reachable belief states sampling. 
Fundamentally, more active antennas will lead
to greater achievable quality of service, lower delay in the SUs, this however associated with higher power consumption. 
\section{System Model}
\begin{figure}
\centering
    \includegraphics[width=0.2\textwidth]{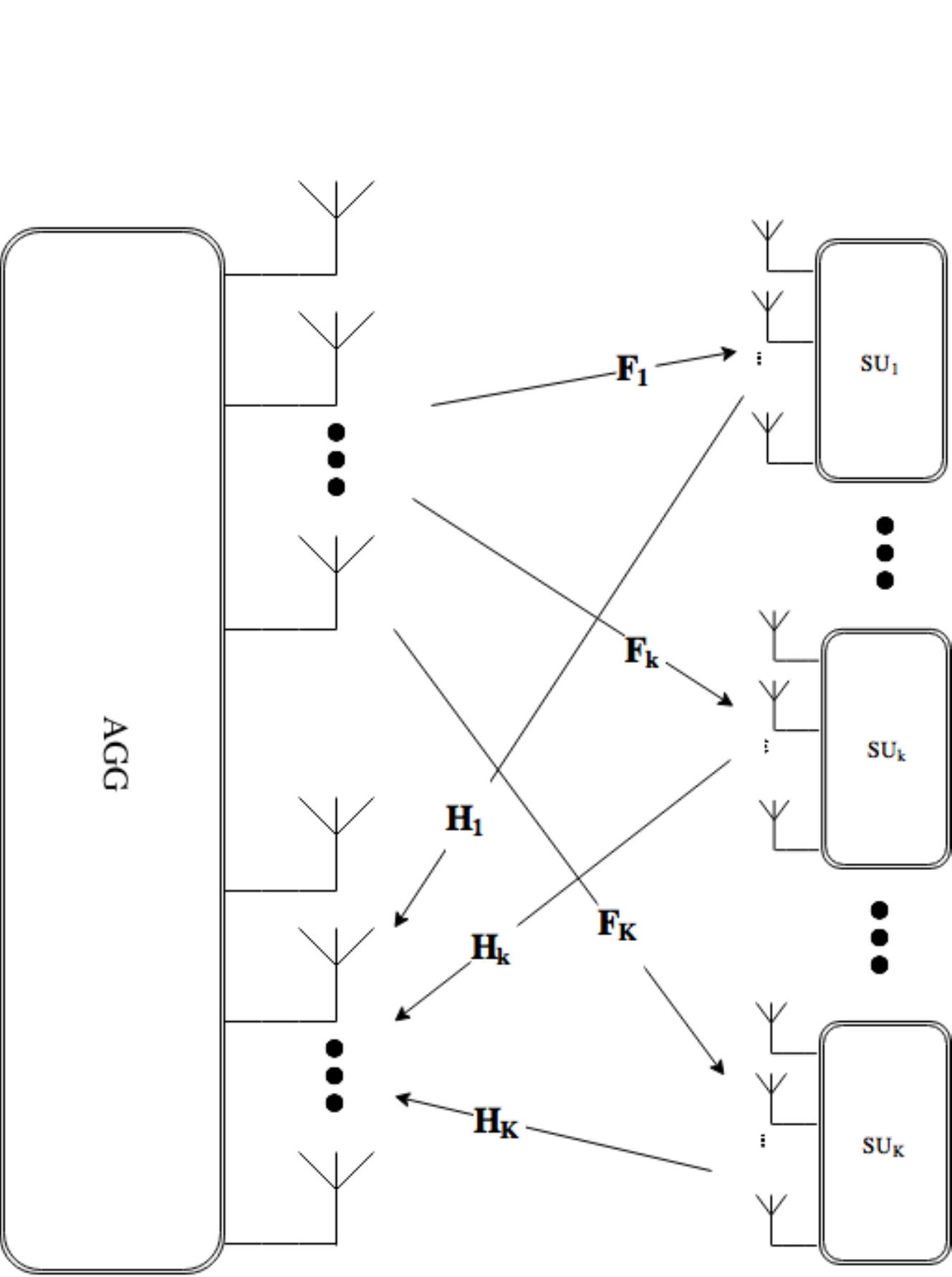}
  \caption{Full duplex mimo system model.}
  \label{fig:FD}
  \vspace{-1em}
\end{figure}

We consider a bidirectional FD SWIPT-MIMO system as shown in fig. \ref{fig:FD}, where $K$ sensor user (SU) indexed with $k \in \mathcal{K} \triangleq \{1, 2, ..., K\}$ and each sensor is equipped with $N_u$ antennas communicate with an aggregator (AGG) equipped with $N_t$ and $N_r$ transmit and receive antennas respectively. Without loss of generality, we assume an equal number of transmit and receive antennas at the AGG and the SUs' are equipped with a small number of antennas compared to the AGG, i.e. $N_t=N_r \gg N_u$. We assume that the AGG is connected to a constant power supply and uncorrelated antennas are assumed at the AGG. The SUs are energy limited devices and harvest their energy from transmitted signal by the AGG. SUs can split the received signal by using power splinter into two different energy harvesting (EH) and information detection (ID) elements. The power splitting (PS) ratio of the $k_{th}$ SU for the EH and ID elements are denoted by $\rho$ and $1-\rho$ respectively. SUs are also equipped with a limited capacity rechargeable battery that stores the harvested energy.
\subsection{Imperfect channel knowledge at the AGG}
 In the realistic wireless communication system, due to the feedback delays and/or estimation error, actual
channel  is different from estimated channel and can be modeled as, 
\begin{align}
\bf H_k=\sqrt{1-\alpha^2}\bf \hat{H}_k+\alpha \bf \Delta_k,
\end{align}
where $\bf V_k^u$ and $\bf w_k^u$ are the antenna selection and beamforming matrices respectively,  $\hat{H}_k\sim \mathcal{C}\mathcal{N}(0,\bf I)$ is the imperfect estimated channel with zero mean and unit variance at the AGG and $\Delta_k \sim \mathcal{C}\mathcal{N}(0, \bf I)$ is the estimated channel Gaussian noise at AGG and $\alpha$ is the channel uncertainty factor. The received signal at the AGG after applying the antenna selection is denoted as,
\begin{align}
\bf{y}_k^u\triangleq&~ {\sqrt{p^u}{(\sqrt{1-\alpha^2}\bf \hat{H}_k+\alpha \bf \Delta_k)\bf V_k^u\bf{w}_k^u\bf{s}_k^u}} \notag \\&+ { {\sqrt{p^u}\sum_{i\neq k, i\in K}{(\sqrt{1-\alpha^2}\bf \hat{H}_i+\alpha \bf \Delta_i) \bf V_i^u\bf{w}_i^u\bf{s}_i^u}}} \notag \\&+ {{\sqrt{p^d} \bf{G}^d \bf V^u\bf{w}^d\bf{s}^d}}+\bf V_k^u\bf{n}_k^u \notag \\ 
&=\sqrt{1-\alpha_k^2} \check{H}_k
\begin{bmatrix}
\bf V_k^u \bf w_k^u \bf s_k^u \\ \\
\sum_{i\neq k, i\in K}{\bf{\hat{C}}_i}\bf{s}_i^u  + \bf \check{\bf G}^d \bf s^d
\end{bmatrix}
+ \notag \\ & \alpha_k \Delta \bf V^u\bf w^u \bf s^u+\bf n_{k,s}^u,
\label{eq:u_y_im}
\end{align}
where $\bf \check{H}_k\triangleq \left[ \bf{\hat{H}}_k  \quad \bf\check{A}_i\right]$, $\bf\check{A}_i \bf\hat{C}_i=\bf\hat{H}_i\bf V_i^u \bf{w}_i^u, \:  \{\forall\: i\in K \: \& \: i\neq k\}$, and $\bf\check A_i \check{\bf G}^d=\bf G^d \bf V^u \bf w^d / \sqrt{1-\alpha_k^2}$ are the uplink and downlink interference coefficient respectively, $\Delta=[\Delta_1,\: \Delta_2,\: .\:.\:.\: \Delta_K]$, $\bf w^u=[\bf w_1^u,\: \bf w_2^u,\: .\:.\:.\: \bf w_K^u]$ and $\bf s^u=[\bf s_1^u,\: \bf s_2^u,\: .\:.\:.\: \bf s_K^u]$. 
The ZF beamforming equalizer at the AGG receiver can be implemented by precode filter \cite{Yuk-FanHo2008} and expressed as,
\begin{align}
\bf \check U_k^{zf}=\left[\bf I_{d_k} \quad \bf 0 \right] \check{\bf H}_k^{-1}=\bf \check{Z}_k\check{\bf H}_k^{-1}.
\label{ZF2}
\end{align}

\begin{prep}
For multi user uplink FD MIMO system with imperfect CSI at AGG and adopting ZF beamforming $\bf \check {U}_k^{zf}$, the processed SINR distribution is as follows:
\begin{align}
f(\gamma_{k,s}^{uzf})=\frac{{\gamma_{k,s}^{uzf}}^{(2N_r\!-\!N_u\!-\!\!1)/2}\!\!exp\!\left[\!-\frac{1}{2}tr(\bf ({\boldsymbol\eta_k}^\mathcal{H}{I_{dk}}{\boldsymbol\eta_k})^{-1}\gamma_{k,s}^u)\!\right]}{(2)^{N_rN_u}\Gamma_{N_{uzf}}(N_r)\det(\bf {{\boldsymbol\eta_k}^\mathcal{H}I_{d_k}{\boldsymbol\eta_k}})^{N_r}},
\label{sinr_zf_i}
\end{align}
where $\boldsymbol\eta_k=\frac{\sqrt{(1-\alpha_k^2)p^u}(  \bf V_k^u \bf w_k^u)}{\sqrt{d_k(\alpha_k^2 p^d \mathcal{J}+\sigma_k^{u})}}$ and $\mathcal{J}=tr({\bf V^u}^ \mathcal{H} {\bf w^u}^\mathcal{H} \bf w^u \bf V^u/d)$. and $d=\sum_i=1^K d_i$, and the uplink SINR is presented in eq.\ref{eq:u_sinr5}
\end{prep}

\subsection{Downlink with imperfect CSI }
Similar to the uplink case, the actual
channel  and the estimated channel are defined as, 
\begin{align}
\bf F_k=\sqrt{1-\alpha^2}\bf \hat{F}_k+\alpha \bf \Delta_k^d,
\end{align}
where similarly, $\hat{F}_k\sim \mathcal{C}\mathcal{N}(0,\bf I)$ is the downlink imperfect estimated channel with zero mean and unit variance and $\Delta_k^d \sim \mathcal{C}\mathcal{N}(0, \bf I)$ is the estimated downlink channel Gaussian noise. The received signal at the $k_{th}$ SU will be as follows,
\begin{align} 
\bf{y}_{k}^{d}\triangleq&~\underbrace {\sqrt{p^d}{(\sqrt{1-\alpha^2}\bf \hat{F}_k+\alpha \bf \Delta_k^d)\bf {V_k^d}\bf{w}_k^d\bf{s}_k^d}} \notag \\[2pt]&+\,\underbrace {\displaystyle \sum _{i\neq k, i \in K} \sqrt{p^d}{(\sqrt{1-\alpha^2}\bf \hat{F}_i+\alpha \bf \Delta_i^d)\bf {V_i^d}\bf{w}_i^d\bf{s}_i^d}}_{{\sf\text { DL interference}}} \notag \\[2pt]&+\,\underbrace {\displaystyle \sqrt{p^u}{\bf{G}^u\bf{w}^u\bf{s}^u}}_{{\sf\text { UL intracell interference}}} + \bf{n}_k^d. 
\end{align}
The downlink SINR is presented in eq.\ref{eq:d_sinr4} and the received ID and EH signal elements are respectively denoted as,

\begin{align}
\textbf{y}_k^{di}=&\sqrt{\rho_k} [ {\sqrt{p^d}{(\sqrt{1-\alpha^2}\bf \hat{F}_k+\alpha \bf \Delta_k^d)\bf {V_k^d}\bf{w}_k^d\bf{s}_k^d}} +\notag \\& {\displaystyle \sum _{i\neq k, i \in K} \sqrt{p^d}{(\sqrt{1-\alpha^2}\bf \hat{F}_i+\alpha \bf \Delta_i^d)\bf {V_i^d}\bf{w}_i^d\bf{s}_i^d}} \notag \\&+\, {\displaystyle \sqrt{p^u}{\bf{G}^u\bf{w}^u\bf{s}^u}}]+\bf{n}_k^s,
\label{eq:d_i}
\\
\textbf{y}_k^{de}=&\sqrt{1-\rho_k}[ {\sqrt{p^d}{(\sqrt{1-\alpha^2}\bf \hat{F}_k+\alpha \bf \Delta_k^d)\bf {V_k^d}\bf{w}_k^d\bf{s}_k^d}} +\notag \\ & {\displaystyle \sum _{i\neq k, i \in K} \sqrt{p^d}{(\sqrt{1-\alpha^2}\bf \hat{F}_i+\alpha \bf \Delta_i^d)\bf {V_i^d}\bf{w}_i^d\bf{s}_i^d}} \notag \\&+\, {\displaystyle \sqrt{p^u}{\bf{G}^u\bf{w}^u\bf{s}^u}}+\textbf{n}_k^d].
\label{eq:d_e}
\end{align}
\begin{prep}
For multi user downlink FD MIMO system with imperfect CSI and the processed SINR follows a matrix variate Beta type $II$ distribution with parameters $(N1,N2)$ and defined as,
\begin{align} 
&\boldsymbol\gamma_k^d\approx B_{N_u}^{II}(N1,N2)\sim \notag \\ &\frac{\det(\boldsymbol \gamma_k^d)^{(\frac{2N1-N_u-1}{2})}\det{(\bf {I_{qk}+\boldsymbol{\gamma}_k^d)}^{{-N1-N2}}}}{\beta( N1,N2)},
\label{gamma_d}
\end{align}
where 
\begin{align}
\beta(N1,N2)= \frac{\Gamma_{N_u}(N1)\Gamma_{N_u}(N2)}{\Gamma_{N_u}({N1+N2})},
\end{align}
and $\Gamma_{N_u}(x)$ is the multivariate Gamma function given as,\vspace{-0.5em}
\begin{align}
\Gamma_{N_u}(x)=\pi^{{N_u}({N_u}-1)/4}\prod_{i=1}^{N_u}\Gamma(x-(i-1)/2).
\end{align}
\end{prep}
By considering, \vspace{-0.5em}
\begin{align}
\boldsymbol\eta_k^g=\frac{K(1+(\frac{p^u}{(1-\alpha^2)p^d})^2)-1}{K(1+\frac{p^u}{(1-\alpha^2)p^d})-1} \\
N_g=\frac{2N_t(\frac{p^u}{(1-\alpha^2)p^d}K+K-1)^2}{(\frac{p^u}{(1-\alpha^2)p^d})^2K+K-1},\\
\boldsymbol\eta_k^q=\frac{N_g{\boldsymbol\eta_k^g}^2+2N_tK(\frac{\alpha^2}{1-\alpha^2})^2}{N_g{\boldsymbol\eta_k^g}+2N_tK\frac{\alpha^2}{1-\alpha^2}}\\
N_q=\frac{(\frac{2N_tK\alpha^2}{(1-\alpha^2)}+N_g\boldsymbol\eta_k^g)^2}{2N_tK(\frac{\alpha^2}{(1-\alpha^2)})^2+N_g{\boldsymbol\eta_k^g}^2},\\
\boldsymbol\eta_k^v=\frac{\boldsymbol\eta_k^q N_q}{N_q+\sigma_k^{d0}},\\
N_v=N_q/2+\frac{\sigma_k^{d0}(2N_q+\sigma_k^{d0})}{2N_q},
\end{align}
according to the definition of  $B_{N_u}^{II}(N1,N2)$ in \cite{Ivrlac2003} by invoking the first and second moment,  $N1$ and $N2$ degrees of freedom are given as follows,
\begin{align}
&N1=\frac{N_t(N_t+(N_v-2)\eta_k^v+1)}{\eta_k^v(N_t+N_v-1)},\\ & N2=\frac{N_v(N_t-3\eta_k^v+2)+N_v^2\eta_k^v+2(\eta_k^v-1)}{N_t+N_v-1},
\end{align}
\newcounter{storeeqcounter4}
\newcounter{tempeqcounter4}
\addtocounter{equation}{0}%
\setcounter{storeeqcounter4}{\value{equation}}%
%
 \begin{figure*}[!t]
 \normalsize
\setcounter{tempeqcounter4}{\value{equation}} 
\setcounter{equation}{\value{storeeqcounter4}} 
\begin{align}
{\gamma}_{k,s}^{u,zf}=\frac{(1-\alpha_k^2)P_k\bf{V_k^u}^\mathcal{H} \bf {w_k^u}^\mathcal{H} \bf{w}_k^u \bf V_k^u/ d_k }{  \bf {Z_k}^\mathcal{H}(tr(\alpha^2  \bf {V^u}^\mathcal{H} {\bf w^u}^\mathcal{H}  P \bf w^u \bf V^u) (\check{\bf{H}}_k^\mathcal{H} \check{\bf H}_k)^{-1}+{\sigma_n^u}^2(\check{\bf{H}}_k^\mathcal{H} \check{\bf H}_k)^{-1}   )\bf{Z}_k},
\label{eq:u_sinr5}
\end{align}
\begin{align}
{\gamma}_k^{d}=\dfrac{{\bf \hat{F}_k^{\mathcal{H}}\bf {V_k^d}^{\mathcal{H}}{\bf{w}_k^d}^{\mathcal{H}}\bf{w}_k^d}\bf {V_k^d}\bf\hat{F}_k}{[{{\sum_{i=1, i\neq k}^{K}{\bf{\hat{F}}_k^{\mathcal{H}}\bf {V_i^d}^{\mathcal{H}}{\bf{w}_i^d}^{\mathcal{H}}\bf{w}_i^d}\bf {V_i^d}\bf{\hat{F}}_k+\frac{\alpha^2}{(1-\alpha^2)}\sum_{i=1}^{K}{\bf{{\Delta}_i^d}^{\mathcal{H}}\bf {V_i^d}^{\mathcal{H}}{\bf{w}_i^d}^{\mathcal{H}}\bf{w}_i^d}\bf {V_i^d}\bf{{\Delta}_i^d}
+{{\frac{p^u}{(1-\alpha^2)p^d}}{\bf{G}^u}^{\mathcal{H}}{\bf{w}^u}^\mathcal{H}\bf{w}^u}\bf{G}^u}}}\notag\\ +\frac{{\sigma_k^d}^2}{(1-\alpha^2){p^d}}+\frac{{\sigma_k^s}^2}{\rho_k(1-\alpha^2){p^d}}]
\label{eq:d_sinr4}
\end{align}
 \hrulefill
  \vspace{-1.5em}
 \end{figure*}
\subsection{Data Source and Data Queue Model}
In this work we consider a simple queuing system with one queue and
a single server. Let $A_k(t)$ and $R_k(t)$ be the inter-arrival rate and service rate at the $k_{th}$ $SU$ at the $t_{th}$ scheduling time slot respectively. Hence, the dynamic of the data queue is given by: 
\begin{align}
\!\!{Q_k}(t\!\!+\!\!1)\!\!=\!\!min\{[Q_k(t)\!\!-\!\!R_k(t)]^+\!\!+\!\!A_k(t),{Q_k}_{max}\},  \forall k \!\!\in\!\! \mathcal{K}, \!\!
\label{eq:d_data}
\end{align}
where $[x]^+\triangleq max\{x,0\}$ and ${Q_k}_{max}$ is the $k_{th}$ SU's queue maximum capacity.


\subsection{Energy Harvesting and Energy Queue Model}
\color{black}
In this paper, the rechargeable battery at SUs is modeled by a limited buffer of energy storage. 
Because of the random nature of wireless EH process, EH are typically presents by stochastic processes. We use the adjacent transition- finite state Markov chain (AT-FSMC) model to represent the energy arrival and the energy consumption separately. We assume that the energy buffer range is uniformly divided into non-overlapping steps with $\Delta E$ quantization factor. Therefore, the energy buffer level is denoted as $\mathcal{E}=\{E_0,E_1,E_2,,..., E_{max}\}$, where $\mathcal{E}$, is the finite set of energy buffer level, and $E_{0}=0$ and $E_{max}$ are the minimum and maximum possible level of the energy buffer. The temporal queue dynamics of the $k_{th}$ SU is given by,
\begin{align}
\!\!{E_k}(t\!\!+\!\!1)\!\!=\!\!min\{max\{E_k(t)\!\!-\!\!A_k(t),0\}\!\!+\!\!B_k(t),{E_k}_{max}\},\!\!
\label{eq:d_ene}
\end{align}
where $A_k(t)$ and $B_k(t) \in \{B_0, B_1,...\}$, are the number of energy units used and energy arrival units in time slot $t$ at the $k_{th}$ user. 

\section{Problem Formulation and Optimization}


Usually, in MDP the agents' task is to determine the best set of action in an imperfect sensing environment and then
perform its set of action depending on the history of its observations. The agents' actions, therefore defined by a set of policy, where usually it is a mapping from a set of observation into action set. In POMDP programming we have no information about the process states during the decision making process for the next action. Therefore, the next actions are determined by only the available information. Basically, all previous  past actions and
observations are needed to determine the optimal actions. The optimal policy for the
POMDP is defined as mapping from the information states to actions, i.e $\boldsymbol \Omega : \mathcal{S}\rightarrow \mathcal{A}$.
In this work we define the global states as the set of states $\bf \mathcal{S} = \{\bf Q, \bf E, \bf H, \bf{\hat H}, \bf F, \bf{\hat F}\}$ and $\bf \mathcal{S}_k = \{\bf Q_k, \bf E_k, \bf H_k, \bf{\hat H}_k,\bf F_k, \bf{\hat F}_k\}$ as the local system states.
A feasible stationary beamforming policy $\Omega =
\{\Omega_{\bf V},\Omega_{w}\}$, is a mapping from the global observed queue, energy channel states and action ${ {\bf{\mathcal{O}}}}= \{\bf{Q},\bf{E}, \hat{\bf{H}}, \hat{\bf{F}}\}$ rather than the global system
state ${\bf{\mathcal{S}}}= \{\bf{Q},\bf{E},\bf{H}, \hat{\bf{H}},\bf{F}, \hat{\bf{F}}\}$  to the beamforming actions, where $\Omega_w=\{\bf w^u, \bf w^d\}$, $\Omega_V=\{\bf V^u, \bf V^d\}$ are the beamforming and antenna selection policy respectively and  $\Omega=[\Omega_1,\:,\Omega_2, \:.\:.\:.\:\Omega_K]$ and $\bf w^u=[\bf w_1^u,\: \bf w_2^u,\: .\:.\:.\: \bf w_K^u]$ and $\bf w^d=[\bf w_1^d,\: \bf w_2^d,\: .\:.\:.\: \bf w_K^d]$ are the total beamforming policy, uplink and downlink bemforming policy for $K$ users respectively.
Assuming the possible stationary beamforming policy $\Omega$, the random process $\bf \mathcal{S} = \{\bf Q, \bf E, \bf H, \bf{\hat H}, \bf F, \bf{\hat F}\}$ is a
controlled Markov chain process where the transition probability can be defined as follows, 
\begin{align}
&\Pr \left[{\bf{{S}}}(t+1)\vert{\bf{{S}}}(t),\Omega\left(\bf Q(t),\bf E(t),\hat{\bf{H}}(t), \hat{\bf{F}}(t)\right)\right]\notag \\ &\quad=\Pr\left[{\hat{\bf{H}}(t+1),\bf H}(t+1)\right] \times \Pr\left[{\hat{\bf{F}}(t+1),\bf F}(t+1)\right]\notag \\ & 
\times\Pr\left[{\bf E}(t+1)\vert{\bf{{S}}}(t),\Omega\left(\bf Q(t),\bf E(t),\hat{\bf{H}}(t), \hat{\bf{F}}(t)\right)\right]\notag \\ & \times\Pr\left[{\bf Q}(t+1)\vert{\bf{{S}}}(t),\Omega\left(\bf Q(t),\bf E(t),\hat{\bf{H}}(t), \hat{\bf{F}}(t)\right)\right].
\end{align}
Therefore, the $K$ users' queue dynamics are correlated via the beamforming control policy $\Omega$.
\section{Cost functions and constraints}
Given the  the queue-aware policy set $\Omega$, and the Markov initial state $\bf S(0)$  the average delay of the $k_{th}$ user can be formulated as follows,\vspace{-0.5em}
\begin{align}
\bar{D}^{\Omega}({S}(0))=\lim_{T \rightarrow \infty}sup \frac{1}{T}\sum_{t=1}^{T} \mathbb{E}^{\Omega}\left[\sum_{k=1}^{K} \varrho f(Q_k(t)) \right],
\label{delay1}
\end{align}
where $\varrho_k$ is a positive weighting factors that shows the $k_{th}$ user delay requirement priority and $f(Q_k(t))$ is an increasing function of the $k_{th}$ user delay. Note that the beamforming policy is constrained by the 
stored harvested energy constraint $A_k(t) \leq E_k(t), \: \forall k\in K $. For simplicity we assume normalized slot duration, therefore the measures of power and  energy become similar and we can use them interchangeably and equivalently. Therefore the average transmission power consumption in uplink and downlink is defined as follows,\vspace{-0.5em}
\begin{align}
\bar{P^u}(\Omega)=\lim_{T \rightarrow \infty}sup \frac{1}{T}\sum_{t=1}^{T} \mathbb{E}^{\Omega}\left[\sum_{k=1}^{K} p^u\| w_k^u\|^2\right],
\label{power1}
\\
\bar{P^d}(\Omega)=\lim_{T \rightarrow \infty}sup \frac{1}{T}\sum_{t=1}^{T} \mathbb{E}^{\Omega}\left[\sum_{k=1}^{K} p^d\|V_k^d w_k^d\|^2\right],
\label{power2}
\end{align}
where $\bar{P}(\Omega)=\bar{P}^u(\Omega)+\bar{P}^d(\Omega)$,  $\bar{P}^u(\Omega)=\sum_{k=1}^K\bar{P}_k^u(\Omega)$ is the total power and $\bar{P}^d(\Omega)=\sum_{k=1}^K\bar{P}_k^d(\Omega)$.
Due to the power-limited users' QoS requirements, the feasible beamforming and antenna selection policy should satisfy the users' average rate constraints as follows,\vspace{-0.5em}
\begin{align}
\bar{R}_{k}^u(\Omega)=\lim_{T \rightarrow \infty}sup \frac{1}{T}\!\!\sum_{t=1}^{T} \mathbb{E}^{\Omega}\left[ R_k^u(t)\right], \forall k \in K,\\
\bar{R}_{k}^d(\Omega)=\lim_{T \rightarrow \infty}sup \frac{1}{T}\sum_{t=1}^{T} \mathbb{E}^{\Omega}\left[ R_k^d(t)\right], \forall k \in K,
\end{align}
and the total achievable rate under the control policy $\Omega$ is defined as, \vspace{-0.5em}
\begin{align}
\bar{R}(\Omega)=\lim_{T \rightarrow \infty}sup \frac{1}{T}\sum_{t=1}^{T} \mathbb{E}^{\Omega}\left[ \sum_{k=1}^K \left(R_k^u(t)+R_k^d(t)\right)\right], 
\end{align}


\begin{prob}
Power-Constrained antenna selection and beamforming Delay optimization Control Policy (DO-CP):
\end{prob}
In this section, we use decision-theoretic POMDP to develop average delay optimization problem. For the initial state, the Qos and energy constrained antenna selection and beamforming problem can be defined as, 
\begin{align}
&\displaystyle \min _{\mathbf {\Omega} \pmb {}} {\mathcal P}_{1}(\mathbf {\Omega} \pmb)\triangleq \bar{D}(\Omega) \\& \mathcal{C}1\quad ~~\displaystyle \bar{P}_k^u(\Omega)\leq P_{k}^{u,max},\quad \forall k\in { K}, \tag{\theequation d} \\&\mathcal{C}2\quad ~~\displaystyle \bar{P}_k^d(\Omega)\leq P_{k}^{d,max},\quad \forall k\in { K}, \tag{\theequation e}\\&\mathcal{C}3\quad ~~\bar{D}_k^u(\Omega) \leq \tau_{k}^u,\quad \forall k\in {K} \tag{\theequation f}\\&\mathcal{C}4\quad ~~\bar{R}_{k}^u(\Omega) \ge r_k^{u,\min }, \quad \forall k\in K, \qquad \qquad ~ \tag{\theequation g}\\&\mathcal{C}5\quad ~~\bar{R}_{k}^d(\Omega) \ge r_k^{ d,\min }, \quad \forall k\in K, \qquad \qquad ~ \tag{\theequation g}
\end{align}
here, $\mathcal{C}1$ and $\mathcal{C}2$ show peruser instantiation power constraints which is determined by the EH process. $\mathcal C3$ gives 
the maximum average delay that the user can tolerate in the
uplink. $\mathcal C4$ specifies the minimum required QoS of each SUs
in the uplink. and finally, $\mathcal C5$ specify the minimum required downlink rate. 
\subsection{Bellman Equation for POMDP}
The optimal beamforming and antenna selection control policy of an infinite-horizon  partially observed information MDP for discounted scenario can be determined similar to the observed MDP and the optimal policy satisfies the equivalent Bellman equation as follows,
\begin{align}
&{\bf{V}^*}\!({S})\!\!=\!\!\min_{A \in \mathcal{A}} \{ \boldsymbol\rho(\boldsymbol\nu,\boldsymbol\varrho, {S}, A)\!\!+\!\!\boldsymbol\varsigma\!\!\sum_{S'\in\mathcal{S}}\!\!\Pr\{S'|{S},A\}{\bf V^*}\!(S')\!\},
\label{bel1}
\end{align}
where, ${\bf{V}}({S})$ is the value function for states ${S}$. $\boldsymbol\rho(\boldsymbol\nu, \boldsymbol\alpha,S, A)$ is the cost function for the constrained POMDP. To deal with the  constrained POMDP, a set of  Lagrange multipliers $\boldsymbol\nu$ are introduced. These multipliers will formulated the constrained POMDP as an unconstrained model. The cost function for  state $S$ and action policy $A$ in POMDP is given as,
\begin{align}
&\bf \boldsymbol\rho(\boldsymbol\nu,\boldsymbol\varrho, S, \boldsymbol, A)= \sum_{k=1}^{K} [ f(S,A)+ \notag \\ &  \nu_k^{p,u} (\bf\bar{P}_k^u(\bf S,\bf A)-P_{k}^{u,max})+\nu_k^{p,d} (\bar{P}_k^d(S,A)-P_{k}^{d,max})\notag \\ & + \bf \nu_k^{r,u}(r_k^{u,\min }-\bar{R}_{k}^u(S,A) )+\bf \nu_k^{r,d}(r_k^{d,\min }-\bar{R}_{k}^d(S,A) ), \notag \\ & +\bf \nu_k^{\tau,d}(\bar{D}_k^u(S,A) - \tau_{k}^u)] \quad \forall \: \bf S \in \mathcal{S} \quad \& \: \forall \: \bf A  \in \mathcal{A},
\end{align}
where, $\bf f(S,A)$ is the objective function in Problem 1,. $\bf \bar{P}_k^u(\bf S,\bf A)=\bar{P}_k^u(\boldsymbol\Omega)$, $\bf \bar{P}_k^d(\bf S,\bf A)=\bar{P}_k^d(\boldsymbol\Omega)$ and $\bf \bar{D}_k^u(S,A)=\bar{D}_k^u(\Omega)$.
Therefore, from the general Bellman equivalent equation at eq.\ref{bel1} the optimal control policy selects the cost-minimizing action and is given by,  
\begin{align}
&\boldsymbol\Omega^*(\hat S)=arg\min_{A\in\mathcal{A}}\{\{ \boldsymbol\rho(\boldsymbol\nu,\boldsymbol\varrho, {S}, A)+\notag \\ & \boldsymbol\varsigma \sum_{S'\in\mathcal{S}}\Pr\{S'|{S},A\}{\bf V^*}(S') \}\}.
\end{align}
Standard value iteration method for POMDPs is generally used to find the infinite horizon optimal control policy $\boldsymbol \Omega^*$ by using a series of finite horizon optimal value functions $V^{0*},V^{1*},...,V^{t*}$. In this method the t-horizon optimal value function approaches the  optimal value function as iteration $t$ approaches to infinity, \vspace{-0.5em}
\begin{align}
\lim_{t \rightarrow \infty } \max_{S\in \mathcal{S}} |V^*(S)-V^{t*}(S)|=0.
\end{align}






Dynamic programming can be used to find the optimal solution of the finite-horizon MDP models and also for finding near-optimal approximations of the value function  for the discounted finite-horizon model. However, in the POMDP the actual system states are only partially observable,   belief states $b$ are used to calculate the value function of the optimal control policy. Belief states are a set of probability distributions over $\mathcal{S}$, i.e. $b \in \mathcal{B}$ and the next-step belief state can be calculated using the update formula based on the Bayes rule as follows, 
\begin{align}
b(S')=\frac{\Pr(O|S',A)}{\Pr(O|A,b)}\sum _{S \in \mathcal{S}}\Pr(S'|S,A)b(S),
\label{blvu}
\end{align}
where $\Pr(O|A,b)=\sum_{S,S'\in \mathcal{S}}\Pr(O|S',A)\Pr(S'|S,A)b(S)$.
Using the belief state, the POMDP value function can be re-written as follows,
\begin{align}
&{\bf{V}^t}({b})\!\!=\!\!\min_{A \in \mathcal{A}} \{ \boldsymbol\rho(\boldsymbol\nu,\boldsymbol\varrho, {b}, A)\!\!+\!\!\boldsymbol\varsigma\!\!\sum_{O\in\mathcal{O}}\!\!\Pr\{O|A,b\}{\bf V^{t-1}}(b) \},
\label{bel2}
\end{align}
where the belief based cost function $\boldsymbol\rho(\boldsymbol\nu,\boldsymbol\varrho, {b},\!\! A)$ is defined as, 
\begin{align}
\boldsymbol\rho(\boldsymbol\nu,\boldsymbol\varrho, {b}, A)=\sum_{S \in \mathcal{S}} b(S)\boldsymbol\rho(\boldsymbol\nu,\boldsymbol\varrho, {S}, A).
\end{align}
The computational complexity of the POMDP is a result of two sets of operations in every iteration step.  First, the computational complexity due to the belief update operation in eq. \ref{blvu}. Second, the computational complexity that arise from the the optimal
action selection operation which requires finding the solution of the control policy function $\Omega$. Another challenges in solving a POMDP is that the complexity of the equivalent linear and convex piecewise function can rapidly increase with the number of iterations. In particular, the size of the set of linear functions determining the POMDP can expand exponentially by only one iteration step. 
 Several approximation method have been developed to reduce the complexity of this POMDP solutions, for example using heuristic estimates to calculate the value function, or updating the value function only for a limited selection belief points. In this work we consider 
the  point-based (PB) value update estimation, which have been widely utilized in the recent advances in POMDPs solutions.
\begin{rema}
Motivation of Two-Layer Control Policy,
\end{rema}
It is clear that the proposed POMDP optimization in Problem 1 involve non-linear functions as well as binary and continuous variables; therefore it is classified as a non-linear  mixed-integer optimization problems. Furthermore, it is intractable to directly tackle the joint optimization of the antenna selection and beamforming policies. However,  we are able to sub-optimally solve the problem for some of the variables and then find the general solution for the remaining variables for any optimization problems \cite{Resende2006,Saki2015}. Consequently, we develop two stage solution of the POMDP optimization for antenna selection and beamforming policy.  Hence, we find the optimal beamforming policy at first stage (inner-layer) under a preset fixed antenna selection policy. Thereafter, we derive the optimal antenna selection policy (outer-layer) based on the results from the inner-layer process to improve utility function.

\begin{algorithm}
\caption{}\label{Alg2}
\begin{algorithmic}[1]
\State Layer 1
\Procedure{HSVI-BeamForming }{}
\ForAll{$A \in \mathcal{A}$}
\State $\boldsymbol\Omega^t_{w} \gets \text{0}$
\State $\bf b \gets \bf b_0$
\State $\boldsymbol\Omega^t_{V} \gets Initial$
\State $\bf \overline{V} \gets Initial$
\State $\bf \underline{V} \gets Initial$
\While {$\overline{\bf V}(b)-\underline{\bf V}(b)\geq{\boldsymbol\epsilon_c}$}
\State EXPLORE$(b_0, t=0)$

\State $V^* \gets V^t$

\State $\bf \boldsymbol \Omega_w^* \gets \boldsymbol \Omega^t_w$.
\State \Return $\bf \boldsymbol \Omega^*_w$

\EndWhile
\EndFor
\EndProcedure
\State Layer 2
\Procedure{HSVI-Antenna Selection }{}
\ForAll{$A \in \mathcal{A}$}
\State $\boldsymbol\Omega^t_{w} \gets \Omega^*_{w} $ From HSVI-Beamforming
\State $\bf b \gets \bf b_0$
\State $\boldsymbol\Omega^t_{V} \gets \text{0}$
\State $\bf \overline{V} \gets Initial$
\State $\bf \underline{V} \gets Initial$
\While {$\overline{\bf V}(b)-\underline{\bf V}(b)\geq{\boldsymbol\epsilon_c}$}
\State EXPLORE$(b_0, t=0)$
\State $V^* \gets V^t$
\State $\bf \boldsymbol \Omega_V^* \gets \boldsymbol \Omega^t_V$.
\State \Return $\bf \boldsymbol (\Omega^*_V,\Omega^*_w)$
\EndWhile
\EndFor
\EndProcedure
\Function{Explore}{b,t}
  \If {$\overline{\bf V}(b)-\underline{\bf V}(b)\geq\frac{\boldsymbol\epsilon_c}{\boldsymbol\varsigma ^{t}}$}
  \State $A^* \gets arg\min_{A \in \mathcal{A}} \{ \sum_{S \in \mathcal{S}} b(S)\boldsymbol\rho(\boldsymbol\nu,\boldsymbol\varrho, {S}, A)+$
$\boldsymbol\varsigma\sum_{O\in\mathcal{O}}\Pr\{O|A,b\}(\min_{\alpha^{t-1} \in \mathcal{V}_t}\sum_{S\in \mathcal{S}}b(S)\alpha^{t-1}(S)) \}$
  \State $O^* \gets arg \max_{O \in \mathcal{O}} (\Pr(O|b,A^*))(\overline{\bf V}(b)-$
  \State $\underline{\bf V}(b)-\frac{\boldsymbol\epsilon_c}{\boldsymbol\varsigma ^{t+1}})  $
  \EndIf
  \State $EXPLORE(b(A^*,O^*),t+1)$
  \State $backup(b)$ then update $\overline{\bf V}$ and $\underline{\bf V}$
  \State \Return 
\EndFunction
\end{algorithmic}
\end{algorithm}
\begin{rema}
Piecewise-Linear and Convex Value Function:
\end{rema}
It has been shown in \cite{Sondik},  that the belief state value function of a POMDP, in both the infinite-horizon and the finite-horizon cases, can
be closely approximated by the upper envelope of a finite set of linear functions, identified as $\alpha$-vectors. Therefore, we model the value function determined over the belief state $b$  at $t$-step employing this expression as,\vspace{-0.5em}
\begin{align}
\label{vect1}
\bf V^t(b)=\max_{\alpha_i \in \mathcal{V}^t}b.\alpha_i^t,
\end{align}
where the set $\mathcal{V}^t$ contains all $t$-step $\alpha_i$-vectors and $(.)$ is the inner product. Therefore, we can easily evaluate the $\alpha$-vector update for any particular belief point $b$ s follows,
\begin{align}
\label{vect2}
\bf \alpha^{t+1}_b=\argmax_{\alpha_i \in \mathcal{V}^{t+1}}b.\alpha_i^{t+1}.
\end{align}
This operation is defined as $backup(b)=\alpha_b^{t+1}$ 
We can transform a POMDP into a “belief state MDP” and the value function can be obtained using the following backup operator,
\begin{align}
&{\bf{V}^t}({b})\leftarrow \min_{A \in \mathcal{A}} \{ \sum_{S \in \mathcal{S}} b(S)\boldsymbol\rho(\boldsymbol\nu,\boldsymbol\varrho, {S}, A)+\notag \\ & \boldsymbol\varsigma\sum_{O\in\mathcal{O}}\Pr\{O|A,b\}(\max_{\alpha^{t-1}_i \in \mathcal{V}^t}\sum_{S'\in \mathcal{S}}b(S')\alpha^{t-1}_i(S')) \} \\= &\min_{A \in \mathcal{A}} \{ \sum_{S \in \mathcal{S}} b(S)\boldsymbol\rho(\boldsymbol\nu,\boldsymbol\varrho, {S}, A)+\\ & \!\!\!\!\boldsymbol\varsigma\!\!\!\sum_{O\in\mathcal{O}}\!\!(\max_{\alpha^{t-1}_i \in \mathcal{V}^t}\!\!\sum_{S'\in\mathcal{S}}\!\!\Pr\{O|S'\!,\!A\}\!\!\sum_{S\in \mathcal{S}}\!\!\Pr\{S'|S,A\}b(S)\alpha^{t-1}_i(S')\!)\!\}\notag\\
&=\min_{A \in \mathcal{A}} \{  b.\boldsymbol\rho(\boldsymbol\nu,\boldsymbol\varrho,A)+\boldsymbol\varsigma\sum_{O\in\mathcal{O}}\!\!(\max_{\alpha^{t-1}_i \in \mathcal{V}^t}b.\xi_i(O,A))\},
\label{bel3}
\end{align}
where 
\begin{align}
\!\!\xi_i(O\!,A\!)\!)\!\!=\!\! \!\!\sum_{S'\in\mathcal{S}}\!\!\Pr\{O|S'\!,\!A\}\!\!\sum_{S\in \mathcal{S}}\!\!\Pr\{S'|S,A\}b(S)\alpha^{t-1}_i(S')\!)
\end{align}
Therefore, we have,
\begin{align}
backup(b)=\argmax_{A \in \mathcal{A}}b.\xi_i(O,A)).
\label{back1}
\end{align}
\subsubsection*{Point-based value iteration}
As it is described, the main source of intractability in solving the POMDP  is the process of finding the optimal action from a set of actions for any feasible belief point. For example, if we use eq.\ref{bel3} to calculate the value function,  we experience a set of functions where the size increases exponentially in every iteration. 
A general method to avoid this intractability is to limit the computing process by considering only a reduced size belief points set. Consequently, the belief backup process in eq.\ref{back1} is limited to a small number of times, resulting in a small number of $\alpha$-vectors which is limited by the size of the belief states set. 

We consider the Stochastic Simulation by Explorative Action heuristic (SSEA) for points sampling. In this method for every action,  only one observation is considered and the belief states are updated by eq.\ref{blvu}. Thereafter,  the farthest away  belief state from $b$ are greedy selected according to following, \vspace{-0.5em}
\begin{align}
\epsilon_b=\max_{b'\in \hat{\mathcal{B}}}\min_{b \in \mathcal{B}}\norm{b-b'},
\label{greed}
\end{align}
where $\mathcal{B}$ is the reachable beliefs set. Therefore,  this algorithm  maximum adds one belief point for every belief.
 To update the value function, since updating the belief points value is based on the successor belief points value eq.\ref{bel3},  the value function
iteration may convergence faster  by updating the successor belief points value before the current value. The point-based Heuristic Search Value Iteration (HSVI) method is primarily developed based on this concept.  Through sustaining the value function upper bound $\overline{\bf V}(b)$ and lower bound $\underline{\bf V}(b)$ at specific belief point ,  the distance between them can be considered as the value function uncertainty.   If the considered points are the successor points of the $b_0$,  subsequently decreasing the upper and lower bounds results directly in decreasing the $b_0$  bounds. The iteration can be assumed converged when this difference reaches the threshold. To define the structure of this concept, HSVI picks the successor belief points in a way that maximize the excess uncertainty as follows, \vspace{-0.5em}
\begin{align}
exc(b,t_d)=\overline{\bf V}(b)-\underline{\bf V}(b)-\frac{\boldsymbol\epsilon_c}{\boldsymbol\varsigma ^{t}},
\end{align}
where $\boldsymbol \epsilon_c$ is a convergence threshold, and $t_d$ is the degree of $b$ (i.e. number of actions from $b_0$ to $b$). Based on this method the optimal control policy in Problem 1 can be obtained by solving an equivalent Bellman equation over a reduced state space, which is summarized in the Algorithm 1.

%

\begin{figure}
\centering
    \includegraphics[width=0.50\textwidth]{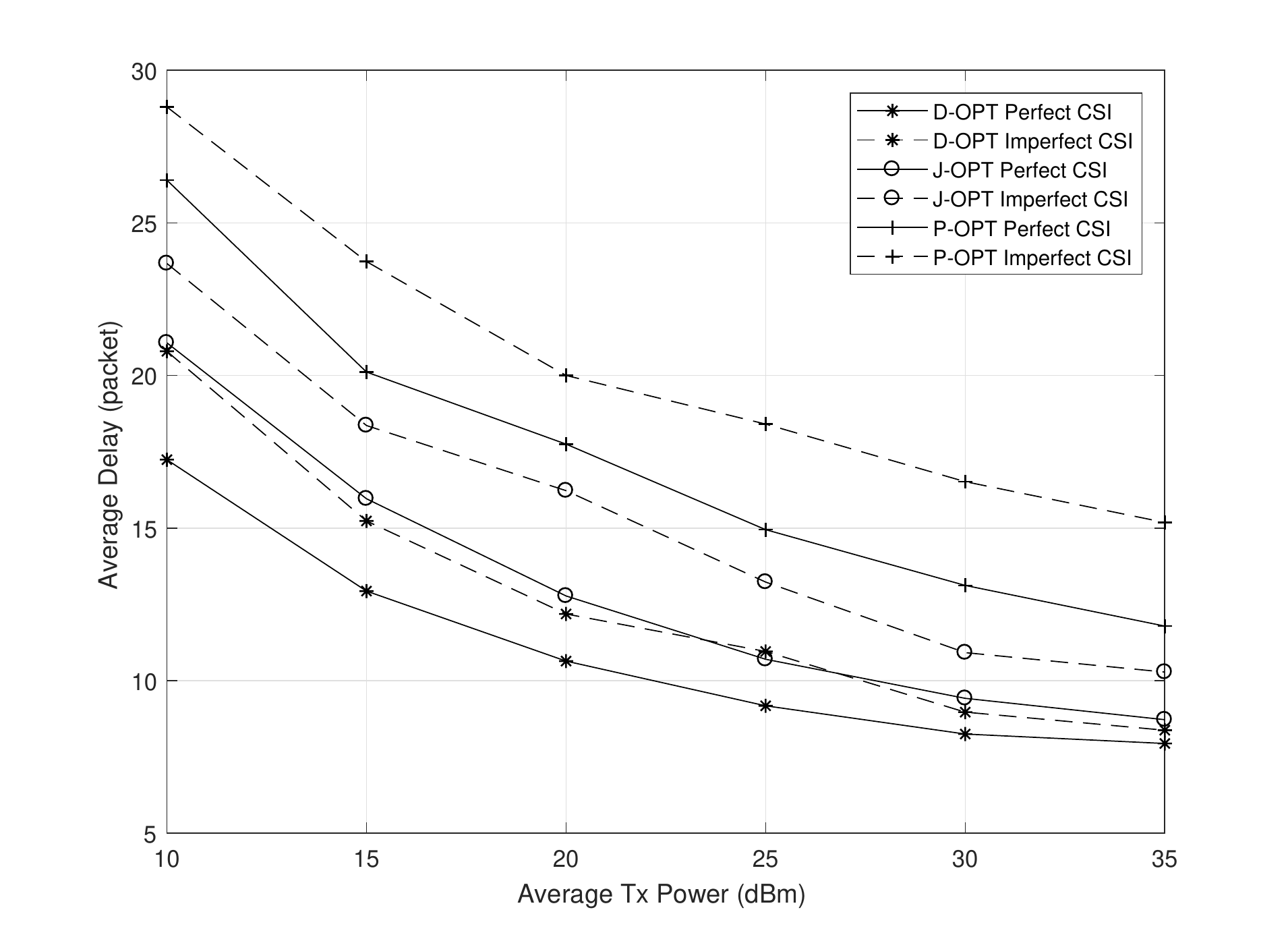}
      \caption{Average delay versus average transmit power for three SUs.}
  \label{fig:6}
\centering
    \includegraphics[width=0.50\textwidth]{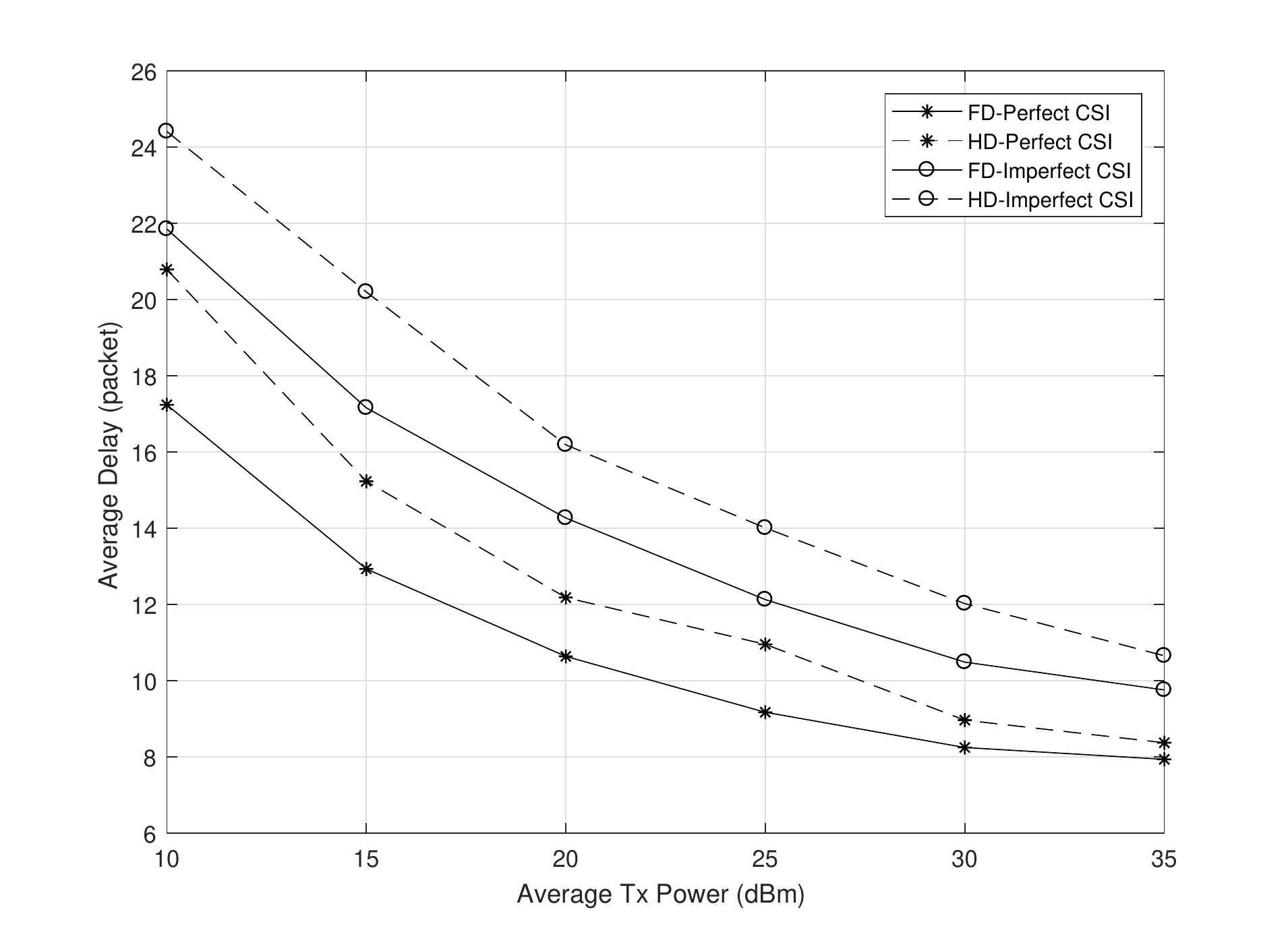}
    \caption{Average delay versus average transmit power for three SUs.}
  \label{fig:7}
   \vspace{-1.5em}
  \end{figure}

\section{SIMULATION RESULTS}
In this section, we study the performance of the proposed delay optimization POMDP method.  To analyze the delay optimization in the SWIPT-MIMO system, we assume that the number of receive and transmit antennas at the AGG is equal and $N_r=N_t=16$. The total bandwidth is 10 MHz. We also consider 3 SUs facilitated with 2 transceiver antennas.
 The packet arrival is considered as a Poisson process with average rate $A_k(t)=10$ (pck/s) and deterministic packet size of 20 Kbits. The decision slot duration is 5 ms and the maximum buffer size is 30 packets. The channel uncertainty is considered as $\alpha=0.2$. The power split factor is assumed to be $\rho=0.5$ and the power efficiency at the EH unit is $\eta_k=40\%$ for all SUs and the harvested energy is stored in a 3.2 v 20Ah battery. 
\begin{figure}[h]
\centering
    \includegraphics[width=0.5\textwidth]{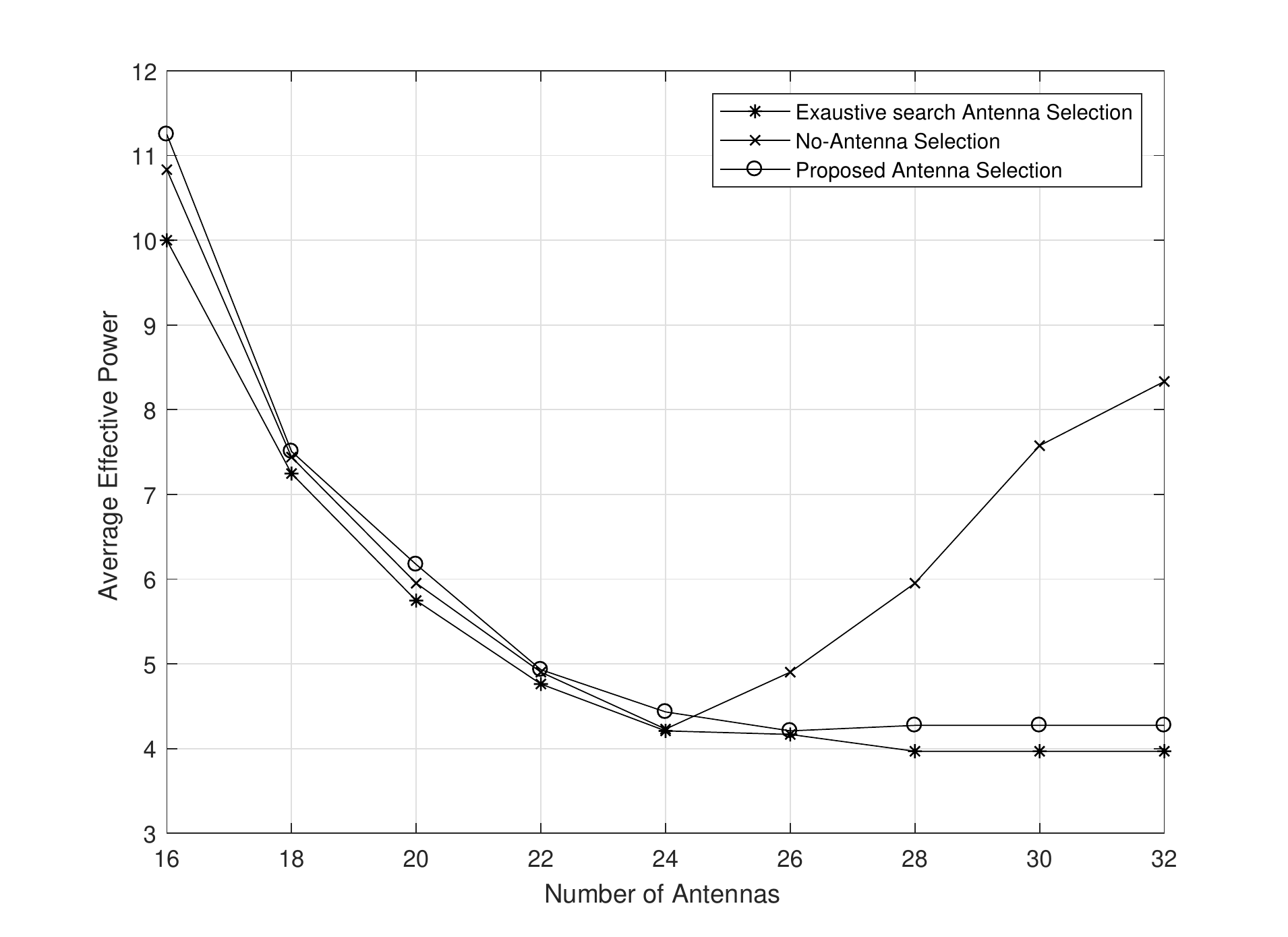}
  \caption{Average effective power versus number of antennas.}
  \label{fig:8}
   \vspace{-1.5em}
\end{figure}
Fig. \ref{fig:7} shows the total average delay for three SUs versus the SUs' total transmit power. The average total delay of the proposed scheme (D-OPT) is compared with the well known joint power and delay optimization (J-OPT) and power optimization (P-OPT) schemes for the cases with perfect and imperfect CSI. It is clear that the average total delay for all schemes improves with at higher transmit power due to better SINR. However, the performance of the D-OPT scheme is considerably better compared to other schemes as a direct result of QoS awareness of the control policy in this scheme. It is observable that the J-OPT is a trade-off between two contradicting objectives i.e. the delay and power consumption. Fig. \ref{fig:7} shows the same parameters for the HD-SWIP-MIMO and FD-SWIPT-MIMO for the perfect and imperfect CSI. It shows that the FD scenario outperforms the HD one. However, at the higher transmission power, the performance of two scenarios tend to converge.  Finally, Fig. \ref{fig:8} shows the average effective transmit power versus the number of receive antennas at the AGG. When $N_r\leq 24$ the antenna selection and no-antenna selection schemes perform similarly. However, for $N_r>24$, the proposed antenna selection scheme considerably outperform the no-antenna selection scheme which confirms the performance of the proposed antenna selection control.
\section{Conclusion}
In this work, we propose a low complexity energy-efficient antenna selection and beamforming control policy to we propose a low complexity optimize the fellow delay for IBFD SWIPT-MIMO systems. We model this optimization problem as a POMDP. We derive a closed-form expression of the value function using belief state value function. Based
on this expression, we developed a conservative formulation of the original POMDP problem and propose an alternating iterative algorithm to efficiently solve the associated problem.  To obtain a low complexity sub-optimal point-based Heuristic Search Value Iteration (PB-HSVI) method is developed. To further reduce the complexity, 
we propose to separate the antenna selection procedure and beamforming operation. Numerical results show that our antenna selection and beamforming control policy significantly perform better compared to the
other methods in the literature.

 
 

\bibliographystyle{IEEEtran}
{\footnotesize
\bibliography{IEEEabrv,Ref_EE_FD}}

\end{document}